\documentclass[conference]{IEEEtran}
\IEEEoverridecommandlockouts
\usepackage{cite}
\usepackage{amsmath,amssymb,amsfonts}
\usepackage{graphicx}
\usepackage{textcomp}
\usepackage{xcolor}
\usepackage{xspace}
\usepackage{mathrsfs}
\usepackage{algorithm}
\usepackage[noend]{algpseudocode}
\usepackage{booktabs}
\usepackage{hyperref}

\def\BibTeX{{\rm B\kern-.05em{\sc i\kern-.025em b}\kern-.08em
    T\kern-.1667em\lower.7ex\hbox{E}\kern-.125emX}}

\newcommand{\Prv}{$\mathcal{P}$\xspace}
\newcommand{\Vrf}{$\mathcal{V}$\xspace}
\newcommand{\Cir}{$\mathcal{C}$\xspace}

\newcommand{\sys}{\texttt{ZKROWNN}\xspace}

\begin{document}

\title{\sys: Zero Knowledge Right of Ownership for Neural Networks}

\author{\IEEEauthorblockN{Nojan Sheybani$^1$, Zahra Ghodsi$^2$, Ritvik Kapila$^1$, Farinaz Koushanfar$^1$}
\IEEEauthorblockA{\textit{$^1$University of California San Diego, $^2$Purdue University} \\
$^1$\{nsheyban, rkapila, fkoushanfar\}@ucsd.edu, $^2$zahra@purdue.edu  }
}

\maketitle

\begin{abstract}


Training contemporary AI models requires investment in procuring learning data and computing resources, making the models intellectual property of the owners.
Popular model watermarking solutions rely on key input triggers for detection; the keys have to be kept private to prevent discovery, forging, and removal of the hidden signatures. 
We present \sys, the first automated end-to-end framework utilizing Zero-Knowledge Proofs (ZKP) that enable an entity to validate their ownership of a model, while preserving the privacy of the watermarks.
\sys permits a third party client to verify model ownership in less than a second, requiring as little as a few KBs of communication. 

\end{abstract}


\section{Introduction}\label{sec:intro}
Deep Neural Networks (DNN) have emerged as the de facto solution for major learning applications such as image and face recognition~\cite{chen2019looks, krizhevsky2017imagenet, hu2015face} and natural language processing~\cite{young2018recent}. Training state-of-the-art DNNs requires access to large amounts of data, as well as massive computational resources; for example, recent language models are trained on terabytes of data, have billions of parameters, and require hundreds of GPUs and algorithmic expertise for training~\cite{lin2021m6, shoeybi2019megatron}.

Given the amount of required data and computing resources spent on training a model, vendors who give access to their trained models remotely or release them publicly have an interest to 
protect the intellectual property rights of their models against copyright infringements. Towards this goal, prior work has proposed methods for watermarking  deep learning models~\cite{darvish2019deepsigns, uchida2017embedding, chen2018deepmarks}.
Watermarks are designed in order to target the decision boundary of the model to resist against various removal attempts such as fine-turning and pruning, while retaining a high accuracy.
Extracting the watermark signature involves providing a special \emph{key input} that triggers the watermark which can be detected at the output, e.g. by a threshold function. However, once the key pertaining to the watermark is revealed, the embedded signature can be easily discovered and removed. This property creates an impediment for litigating an ownership dispute which would require providing proofs of ownership during the discover process to potentially multiple parties. 

In this work, we propose use of Zero-Knowledge Proofs (ZKP) to facilitate legitimate proof of DNN ownership without revealing any other information. ZKPs are a particular set of protocols in cryptography involving a prover (\Prv) and a verifier (\Vrf). The prover seeks to prove a mathematical assertion on a private input to the verifier, without disclosing any other information about the input. While the use of ZKPs in legal settings have been proposed in prior work for other applications ~\cite{bitan2022using, bamberger2022verification}, to the best of our knowledge, our work is the first to propose a concrete framework for DNN proof of ownership. Our work demonstrates how ZKPs can be used by expert witnesses and other parties to verify ownership claims without revealing details on the watermarking procedure that could jeopardize the intellectual property rights of the model owner. The use of ZKPs in legal settings should satisfy two main requirements. First, executing the protocol should be simple, and it is beneficial that the interaction be limited to \Prv sending a single message to \Vrf for verification. Second, the proof has to be \emph{publicly verifiable}, i.e., \Prv can perform the proof generation once that can be verified by any party, without having to convince every entity in a separate process~\cite{bamberger2022verification}.  

Non-interactive Zero-Knowledge Proof systems (NIZK) \cite{blum1991noninteractive} support these requirements. NIZK include a one-time setup, and the process of verifying the proof consists of a single message sent from the prover to verifier. This is in contrast to interactive proof protocols where verification is performed over multiple rounds, and has to be repeated with every new verifier. Our work builds on zero-knowledge succinct non-interactive arguments of knowledge (zkSNARKs) to implement watermark extraction including feed-forward computations in DNNs and sigmoid thresholding for trigger detection. zkSNARK setup and proof generation steps are circuit dependent; if the circuit changes often, this technique could be quite computationally intensive. Fortunately, our proposed work only handles a constant circuit representing the pertinent DNN. Our concrete implementation demonstrates the applicability of our framework, and we show that our framework is able to prove ownership with as little as $11s$ and $1ms$ of computation time on the prover and verifier side respectively, and as little as $35KB$ of communication for image recognition benchmarks. Our setting only requires setup and proof generation once as the circuit does not change, resulting in amortized prover and setup computation time compared to the overall usage lifetime.

In summary, our contributions are:

\begin{itemize}

\item We propose \sys, the first end-to-end watermark extraction and verification framework for deep neural networks based on zero-knowledge proofs. \sys enables a model owner to prove their right of ownership without revealing details of the watermarking technique.

\item \sys incorporates non-interactive proofs for simplified proof generation and verification. More specifically, \sys's proofs are \emph{publicly verifiable}, i.e., the proof is generated once and can be verified by third parties without further interaction.


\item We provide a concrete implementation of \sys with extensive evaluation on various benchmarks. \sys requires small communication (less than 16MB during setup and only 128B to transfer the proof for our largest example), and enables sub-second proof verification.

\end{itemize}


\begin{figure*}[]
    \centering
    \includegraphics[width=\linewidth]{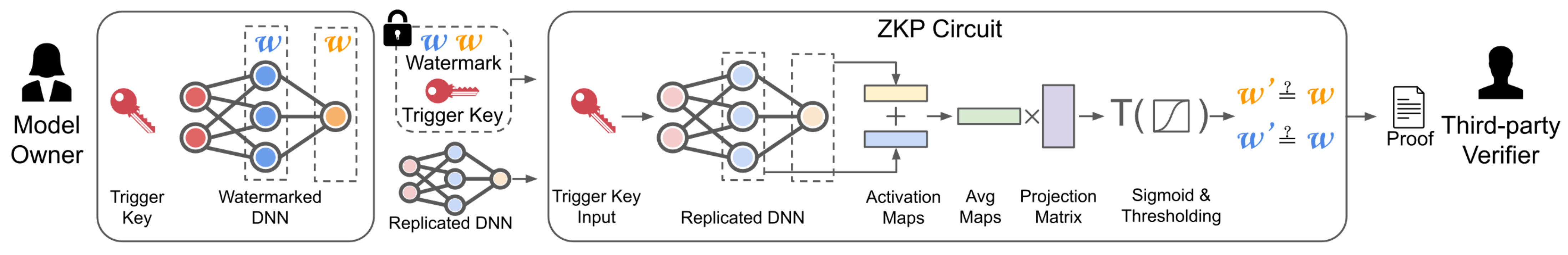}
    \caption{High level description of \sys}
    \label{fig:highlevel}
\end{figure*}

\section{Background}\label{sec:prelims}
\subsection{Neural Network Watermarking} \label{sec:watermarking}
Watermarking techniques have been extended to DNNs to protect the intellectual property of model owners. Watermarking can be considered analogous to introducing a backdoor in a neural network, especially in the case of Black box dynamic watermarking ~\cite{Li2021ASO}. A backdoor is embedded in the model by hand-crafting key input triggers which generate the desired watermark (WM). The model continues to perform at the same accuracy with minimal overhead, except when these selected key inputs are used to verify the watermark. 

Watermarks can be embedded in weights~\cite{nagai2018digital}, activations~\cite{darvish2019deepsigns}, or near the decision boundary~\cite{adi2018turning} of neural networks. 
For image processing networks, ~\cite{Zhang2020ModelWF} proposes spatially invisible watermarking mechanisms. A unified and invisible signature is learned and embedded into all the outputs, which can be extracted afterwards to verify ownership. Additional methods include introduction of statistical bias in the weights while training a neural network~\cite{uchida2017embedding}. An embedding regularizer, which uses binary cross entropy loss, is used in the standard cost function of one of the convolution layers. The watermark can be extracted by projecting w using a secret key X, where the $j^{th}$ bit of the watermark is extracted as $b_j = s(\Sigma_{i}X_{ji}w_{i})$. Here, $s(x)$ is a step function, $w_{i}$ are the weights of the network and X is the secret key required to embed and detect the watermarks. This methodology has advantages over the usual procedure of embedding the signature in the weights of a trained network as it does not degrade the network’s performance after training and embeds the signature during training itself. However, embedding the signatures in the weights of the DNN, even while training, poses significant challenges related to WM robustness, and makes it prone to watermark overwriting and network morphism.

In this work, we consider the watermarking method presented in DeepSigns~\cite{darvish2019deepsigns} which embeds the watermark into the probability density function (PDF) of activation maps across various layers of the network. DeepSigns takes the trained model along with an owner-defined watermark signature, and produces a watermarked network. 

DeepSigns watermark embedding is a two step process, beginning with securely generating owner-specific WM keys. In the next step, the owner’s DNN is fine tuned and the generated WM signature is embedded into the pdf distribution of the activation maps of selected layers. The encoded watermark signatures are Independently and Identically Distributed (iid) arbitrary binary strings. For the intermediate hidden layers, the data distribution is assumed to be a Gaussian Mixture Model (GMM). One or more random indices are selected from 1 to $S$, where each index corresponds to a Gaussian in the mixture model. $S$ is the number of classes in the final application. The WM signature is then embedded into the mean of these selected Gaussian distributions.

The WM keys contain three parameters, the chosen Gaussian classes $s$, the input triggers, which are basically a subset (1\%) of the input training data ($X^{key}$), and the projection matrix $A$. The projection matrix is used to project the mean values of the selected Gaussian distributions into binary space. To minimize the distance between these mean values, that is the centers of the Gaussian distributions and the owner’s WM signature, an additional loss term is added to the cost function while fine tuning. 

The watermark extraction phase consists of three steps. It begins with querying the underlying DNN with the owner-specific watermark keys ($X^{key}$) generated during embedding. In the next step, the Gaussian Centers are approximated by taking a statistical mean of the obtained activation maps. These Gaussian centers and the projection matrix $A$, obtained from the owner’s WM keys, are used to estimate the relevant Watermark signature. Finally, the bit error rate (BER) between the obtained WM signature and the owner's actual signature is computed. If the BER is zero for any layer, DeepSigns ascertains that the deployed DNN is the IP of the model owner in question. This WM methodology is robust to watermark overwriting, model fine-tuning and model-pruning.


\subsection{Zero-Knowledge Proofs}
ZKPs are a cryptographic primitive that allows a prover \Prv to convince a verifier \Vrf that an evaluation of computation \Cir on \Prv's private input $w$, also called the witness, is correct without revealing anything about $w$. 
In a standard ZKP scheme, \Prv convinces \Vrf that $w$ is a valid input such that $y=\mathcal{C}(x, w)$, in which $x$ and $y$ are public inputs and outputs, respectively. When the communication of this proof is done in a single message, the ZK scheme is referred to as non-interactive. ZKPs can also be generated interactively, in which the proof is computed through several rounds of communication between \Prv and \Vrf, but this requires \Vrf to be online for the duration of proof generation, which is undesirable when there are many verifiers. Interactive ZKP schemes are limited, as they only support the \textit{designated verifier} model, meaning that a new proof must be generated for each verifier for one circuit $C$.
In contemporary ZK constructions, \Cir is expressed as an efficient generalization of an arithmetic circuit, such as Rank 1 Constraint Systems (R1CS) or Quadratic Arithmetic Programs (QAPs), which have been popularized due to their ease of use \cite{ben2019aurora, danezis2014square}. 

Zero-knowledge succinct non-interactive arguments of knowledge (zkSNARKs) have been utilized for a myriad of tasks, including the real-world case of Zcash's privacy preserving cryptocurrency \cite{sasson2014zerocash}. zkSNARKs have emerged as a popular ZKP method, acting as the technical foundation for many ZK works, as this construction provides fast and computationally inexpensive proof verification \cite{maller2019sonic}. zkSNARKs also benefit from being \textit{publicly verifiable}, meaning that any verifier with the proper verification key can verify a zkSNARK. The main drawback of zkSNARKs are the reliance on a trusted setup for every new circuit \Cir 
and the intensive computation necessary for proof generation. If \Cir does not change often, or at all, these computational drawbacks can be amortized.


In this work, we use the Groth16 zkSNARK protocol, which is based on QAP representations of computation \cite{groth2016size}. The high-level approach for proof generation with Groth16 (and other NIZKs in general) can be represented with the three following algorithms:

\begin{itemize}
    \item $(\mathcal{VK, PK})\xleftarrow[]{}$ Setup(\Cir): A trusted third party or \Vrf run a setup procedure to generate a prover key $\mathcal{PK}$ and verifier key $\mathcal{VK}$. These keys are used for proof generation and verification, respectively. This setup must be repeated each time \Cir changes.
    \item $\pi \xleftarrow[]{}$ Prove($\mathcal{PK}$, \Cir, $x$, $y$, $w$): \Prv generates proof $\pi$ to convince \Vrf that $w$ is a valid witness.
    \item $1/0 \xleftarrow[]{}$ Verify($\mathcal{VK}$, \Cir, $x$, $y$, $\pi$): \Vrf accepts or rejects proof $\pi$. Due to soundness property of zkSNARKs, \Vrf cannot be convinced that $w$ is a valid witness by a cheating \Prv.
\end{itemize}

The Groth16 protocol allows us to achieve small proofs and fast verification, independent of the circuit size \cite{groth2016size}, at the cost of high prover and setup complexity. Due to the static nature of \Cir in our \sys, proof generation and setup only happen once, so runtimes are amortized and therefore negligible.
\section{Methodology}
\subsection{\sys Setting and Threat Model}

In this work we assume a setting where a model owner holds a watermarked model $\mathbf{M}$ with private trigger key $\mathcal{K}$ and watermark parameters $\mathcal{W}$. The model owner claims that a \emph{second} model $\mathbf{M'}$ is built based on watermarked model $\mathbf{M}$.
The model owner takes the role of a prover \Prv, and generates a proof $\pi$ attesting that $\mathbf{M'}$ produces the watermark $\mathcal{W}$ when triggered with $\mathcal{K}$. In our threat model, prover \Prv is semi-honest, meaning that \Prv will not deviate from the protocol. The ownership proof $\pi$ can be verified by any third party \Vrf, requiring only a verification key. The proof generation and verification steps in \sys are illustrated in Figure~\ref{alg:extract}.
\sys utilizes zkSNARKs to enable proof of ownership without revealing the trigger key $\mathcal{K}$ and watermark $\mathcal{W}$. We extend the watermark embedding and extraction technique in DeepSigns~\cite{darvish2019deepsigns}. As detailed in Section~\ref{sec:watermarking},
%
%
the watermark is embedded in a specific layer, which is only known to the original model owner. This watermark is only extractable when the model takes in a specific trigger key as an input.
As discussed in \cite{darvish2019deepsigns}, the watermarks are embedded in and extracted from the probability density function (pdf) of the activation maps in the model. With all of this information in hand, we outline \sys's zero-knowledge watermark extraction in algorithm \ref{alg:extract}.

\begin{algorithm}[H]
\small
\caption{\sys Watermark Extraction}
\label{alg:extract}
\begin{algorithmic}
    \State {\bfseries Public Values:} Model $\mathcal{M}$, target BER $\theta$
    \State {\bfseries Private Input:} trigger key $X^{key}$, $B$-bit watermark $wm$, projection matrix $A^{M \times N}$ where $M= $ size of feature space and $N= $ size of $wm$, embedded layer $l_{wm}$
    \State {\bfseries Circuit:} 
    \State \hspace{1em} $check=1$
    \State \hspace{1em} $zkFeedForward(\mathcal{M})$ on input $X^{key}$ until layer $l_{wm}$
    \State \hspace{1em} Extract activation maps ${a}$ at layer $l_{wm}$
    \State \hspace{1em}  $\mu^{1 \times M}=  zkAverage(a) $
    \State \hspace{1em} $G^{1 \times N} = zkSigmoid(\mu^{1 \times M} \times A^{M \times N})$
    \State \hspace{1em} $\hat{wm} = zkHard\_Thresholding(G^{1 \times N}, 0.5)$
    \State \hspace{1em} $valid\_BER = zkBER(wm, \hat{wm}, \theta)$ 
    \State \hspace*{1em}  \textbf{return} $check \land valid\_BER$
    
\end{algorithmic}
\end{algorithm}

To support the circuit presented in algorithm \ref{alg:extract}, we provide seperate smaller zkSNARK circuits for each computation, such as sigmoid and thresholding. For the feed-forward operation, we support Dense, ReLU, and Convolution3d layers, as we assume that the watermarks are embedded in one of the initial layers of the model. In addition, we provide end-to-end zero-knowledge watermark extraction circuits for a Multilayer Perceptron (MLP) and Convolutional Neural Network (CNN). In the next section, we provide the implementation details of each operation that \sys supports.



\subsection{\sys Implementation}
To implement \sys, we use \texttt{xJsnark} \cite{xjsnark}, a high-level framework that enables zkSNARK circuit development. The generated circuits are compiled in \texttt{libsnark}, a C++ framework for zkSNARK generation \cite{libsnark}.  As stated before, proof generation and verification are done with the Groth16 protocol using the BN128 elliptic curve, which provides 128 bits of security. While \texttt{xJsnark} and \texttt{libsnark} have open-source gadgets/arithmetic circuits available for general computation, none were relevant to the computation that \sys requires. Therefore, all circuits were designed specifically for watermark extraction, however can be generalized and extended for other relevant applications, such as DNN inference.

zkSNARK arithmetic circuits do not natively support floating point computation without requiring conversion to binary circuits \cite{garg2022succinct}. This process incurs large overhead for the prover, which is already the computational bottleneck in zkSNARK schemes. To avoid floating point computation, we scale our inputs by several orders of magnitude and truncate the result. This does not affect the performance, as the floating point conversions are all done in a preprocessing step before the proof generation and all functions are modified accordingly before circuit generation. For readability, we still use floating point values in our descriptions of algorithm \ref{alg:extract} and our implementations. We now present the implementation details of the functions that \texttt{ZKROWNN} supports. It is important to note that, although these operations are used collectively for end-to-end watermark extraction, each circuit can also be used in a standalone zkSNARK due to our modular design approach.

\subsubsection{Matrix Multiplication}
We implement a zero knowledge matrix multiplication circuit that efficiently computes $A^{M \times N} \times B^{N \times L}=C^{M \times L}$, where $C$ is a private matrix and $A$ or $B$ can be public or private, depending on the application. This circuit can be used for dense/fully connected layers that are done in the feed-forward step of watermark extraction, or for standard matrix multiplication, both of which are necessary in our end-to-end implementation. 



While there have been optimizations for matrix multiplication in zero knowledge proposed before, such as Freivald's algorithm \cite{weng2021mystique}, the most notable optimizations require interactivity between prover and verifier. As the \sys use case greatly benefits from non-interactivity, we do not consider these optimizations.

\subsubsection{Convolution}
We implement the 3D convolution operation by flattening the input and kernel into 1D vectors. The input is grouped and structured based on the size of the kernel and stride value into a vector. Afterwards, we perform a 1D convolution operation between the processed input vector, and the flattened kernel. We develop an arithmetic circuit for zero-knowledge 1D convolution, which consists of inner product and shift operations.


\subsubsection{Sigmoid}
The standard sigmoid function is defined as $S(x)=1/(1+e^{(-x)})$, which is a very difficult computation to do in zero-knowledge. To work around this, we use the Chebyshev polynomial approximation of the sigmoid function presented in \cite{wan2022zk}:

\begin{align*}
    S(x) &= 0.5 + 0.2159198015x - .0082176259x^3  \\
    &+ 0.0001825597x^5 - 0.0000018848x^7 \\
    &+ 0.0000000072x^9
\end{align*}

We reiterate that floating point computation is converted to integer arithmetic by increasing floating point numbers by several order of magnitudes and truncating.

\subsubsection{ReLU and Hard Thresholding}

We implement ReLU in a zero-knowledge circuit that computes $f(x)=max(0,x)$. Due to the similarity between ReLU and hard thresholding, a similar circuit is used for the two operations. To implement hard thresholding, we take in a threshold $\beta$ as an input and build a circuit that computes the following piecewise function:

\begin{equation*}
    f(x)=
    \begin{cases}
        1 & \text{if } x \geq \beta\\
        0 & \text{if } x < \beta
    \end{cases}
\end{equation*}

Hard thresholding is performed on the output of the sigmoid function, resulting in a vector of ones and zeroes that can be concatenated to generate the extracted watermark.

\subsubsection{Bit Error Rate}
The bit error rate is defined as the percentage of bits that differ between the private watermark $wm$ and the \sys extracted watermark $\hat{wm}$. This is the last computation that is done in our end-to-end implementations. To compute this, we perform bit by bit comparison of $wm$ and $\hat{wm}$ as a secondary function implemented in the hard thresholding circuit. If the bit error rate is below some predefined threshold $\theta$, the circuit outputs a 1. If not, the circuit will output a 0.

\subsubsection{End-to-end Examples}
We include implementations of \sys applied to a multilayer perceptron (MLP) and convolutional neural network, assuming that the watermark is embedded in the first hidden layer for both examples. \sys still works when the watermark is embedded in deeper layers, at the cost of higher prover complexity. 


\renewcommand{\arraystretch}{1.3}

\begin{table*}[t]

  \caption{\sys performance benchmarks on all individual zkSNARK circuits and end-to-end examples. All individual, meaning not end-to-end, circuits (e.g. MatMult) are run with private inputs and public outputs, for sake of consistency. 2D operations are run with $128 \times 128$ inputs and 1D operations are run with length $128$ inputs. $Conv3D$ is run with $32 \times 32 \times 3$ inputs with $32$ output channels, $3 \times 3$ filter size, and stride $2$.}
    \vspace{-2mm}
  \centering
  \begin{tabular}{{l}{c}*{6}{c}}
    \toprule
    \textbf{Benchmark} & \textbf{\# Constraints} & \textbf{Setup Runtime (s)} & \textbf{$\mathcal{PK}$ size (MB)}& \textbf{\Prv Runtime (s)} & \textbf{Proof Size (B)} & \textbf{$\mathcal{VK}$ size (KB)} & \textbf{\Vrf Runtime (ms)}\\
    \midrule
    \textbf{MatMult}& 1,097,344 & 57.3976 & 215.6518 & 18.6805 & \textbf{127.375} & \textbf{0.199} & \textbf{0.6}\\
    
    \textbf{Conv3D}& 235,899 & 13.3621 & 46.3793 & 4.2081 & \textbf{127.375} & \textbf{0.199} & \textbf{0.6}\\
    
    \textbf{ReLU}& 8,832 & 0.6384 & 1.7193 & 0.1907 & \textbf{127.375} & \textbf{5.303} & \textbf{0.7}\\
    
    \textbf{Average2D}& 545,793 & 29.6248 & 107.3271 & 9.5570 & \textbf{127.375} & \textbf{5.303} & \textbf{0.6}\\
    
    \textbf{Sigmoid}& 454,656 & 34.4989 & 90.5934 & 8.3680 & \textbf{127.375} & \textbf{41.031} & \textbf{0.8}\\
    
    \textbf{HardThresholding} & 8,704 & 0.624 & 1.6978 & 0.1857  & \textbf{127.375} & \textbf{5.303} & \textbf{0.7}\\
    
    \textbf{BER}& 8,832 & 0.6423 & 1.7526715 & 0.1826 & \textbf{127.375} & \textbf{0.2389} & \textbf{0.6}\\
    \midrule
    \textbf{MNIST-MLP}& 2,093,648 & 68.4456 & 280.3859 & 45.1208 & \textbf{127.375} & \textbf{16,006.343} & \textbf{29.4}\\
    
    \textbf{CIFAR10-CNN}& 590,624 & 32.35 & 117.1699 & 11.22 & \textbf{127.375} & \textbf{34.651} & \textbf{1}\\
    \bottomrule
  \end{tabular}

  \label{tab:perf}
\end{table*}

\section{Evaluation}



\textbf{Experimental Setup.} \sys is implemented with a \texttt{libsnark} \cite{libsnark} backend using the Groth16 zkSNARK protocol. All zkSNARK circuits are built using \texttt{xJsnark} \cite{xjsnark}.  We run all experiments on a 128GB RAM, AMD Ryzen 3990X CPU desktop.

\textbf{\sys Evaluation Metrics.}
We present the following metrics to evaluate \sys and the individual circuits:
\begin{itemize}
    \item \textit{Number of Constraints:} The number of constraints is used as an indicator for how large the zkSNARK circuit is. As number of constraints increases, runtimes also increase.
    \item \textit{Setup Runtime:} The setup process is used to generated the prover key $\mathcal{PK}$ and verifier key $\mathcal{VK}$ by a trusted third party. Trusted setup is a core idea in zkSNARKs. In \sys's setting, this process is only run once, so its runtime can be amortized.
    \item \textit{Prover Runtime:} This is defined as the amount of time for prover \Prv to generate a zkSNARK. As zkSNARKs are designed to reduce verifier complexity, this often comes at the cost of increased prover complexity. Much like the setup, this process is only run once in the \sys setting, so its runtime can be amortized.
    \item \textit{Proof Size:} Due to the succinctness property of zkSNARKs, the proofs that are generated in \sys are very small, requiring very little communication between the prover and all verifiers.
    \item \textit{Prover Key Size:} The prover key size grows with respect to the size witness data in our zkSNARK circuit, so this can grow quite large in our setting. This requires communication from the trusted setup provider to the prover, but, again, this process is only done once.
    \item \textit{Verifier Key Size:} The verifier key size grows with respect to the size of the public inputs in our zkSNARK circuit. This requires communication between the trusted setup provider and each verifier.
    \item \textit{Verifier Runtime:} zkSNARKs aim to minimize verifier complexity, so verifier runtime is often in the millisecond range. This greatly benefits \sys, as our goal is to provide verifiers with a simple scheme to validate their ownership of a model.

\end{itemize}

\subsection{\sys Performance}

\begin{table}[]
\centering
\caption{Details of DNN benchmarks. FC(a) represents a fully connected layer with neurons, and C(a,b,c) represents a convolution layer with output channels, filter size b, and stride c. MP(a,b) represents a max pooling layer with filter size a and stride b.}
    \resizebox{.45\textwidth}{!}{
\begin{tabular}{ll} \toprule
\textbf{Dataset} & \textbf{Architecture} \\ \midrule
\textbf{MNIST} & 784 - FC(512) - FC(512) - FC(10) \\
\textbf{CIFAR10} & \begin{tabular}[c]{@{}l@{}}3$\times$32$\times$32 - C(32, 3, 2) - C(32, 3, 1) - MP(2, 1)\\ C(64, 3, 1) - C(64, 3, 1) - MP(2, 1) - FC(512) - FC(10)\end{tabular} \\ \bottomrule
\end{tabular}}
\label{tab:archs}
\end{table}

We evaluate \sys on two DNN benchmarks: a multilayer perceptron (MLP) on the MNIST dataset and a convolutional neural network (CNN) on the CIFAR-10 dataset.  These benchmarks are extended from DeepSigns \cite{darvish2019deepsigns}. We assume that the model owner embedded a 32-bit watermark in the first hidden layer, however, our framework can handle extracting the watermark from any layer. We also benchmark the specific circuits that make \sys's automated end-to-end framework. \sys does not result in any lapses in model accuracy, as our scheme does not modify the weights of the model at all. \sys is able to achieve the same BER and detection success from extracted watermarks as DeepSigns, while protecting the model owner's trigger keys and preserving the privacy of the watermark. 

Table \ref{tab:perf} highlights the end-to-end performance of \sys on the benchmark architectures described in Table \ref{tab:archs}. The DNN benchmarks use ReLU as the activation function, however we provide the capability of using sigmoid, at the cost of potentially lower model accuracy. Alongside end-to-end performance, we also benchmark the performance of our individual zkSNARK circuits.


When observing the results of \sys, we are able to achieve low communication and runtime for the verifier, even with large circuits. The corresponding results are bolded in Table \ref{tab:perf}. Although we witness relatively high prover/setup runtimes, we reiterate that proof generation and setup only happen once per circuit. In our setting, we benefit from this, as the zkSNARK circuit does not change, thus amortizing the proof generation and setup runtimes and communications. 

Our largest circuit, the MLP circuit, only results in a $127B$ proof. This only requires $29.4ms$ to verify, and any third party with the verifier key can verify this. The verifier key requires $16MB$ of communication from the trusted setup provider to each verifier, due to taking in the model's weights as a public input. Due to memory constraints, we precompute a small portion of the first layer matrix multiplication in the MLP, but ensure that there is no risk of information leakage, as the precomputed values still act as private inputs. The CNN circuit, requiring only a quarter of the constraints as the MLP circuit, has much more desirable setup, prover, and verifier performance. Prover and setup runtimes and proving key sizes are cut at least in half. We are able to maintain the same proof size, with a drastically reduced verifier key, due to the reduction of public input size. This results in a $1ms$ verification time, which is highly attractive for verifiers.

When looking at the results as a whole, we see that proof size stays constant, no matter what the size of the circuit is, which is beneficial in our use case. With our largest individual circuit, matrix multiplication with $128\times128$ inputs, we only need $0.6ms$ to verify computational correctness. As mentioned before, the verifier key grows with the public input, which has a direct effect on the the verification runtime. Some circuits, such as sigmoid and averaging, required some extra public inputs to compute correctness, thus leading to some higher $\mathcal{VK}$ sizes. To reduce runtimes and constraints in our end-to-end example, which are combinations of the individual circuits, we make specific optimizations such as bitwidth scaling between operations and combining operations within loops.


Overall, we show the efficiency of \sys in developing proofs of model ownership alongside fast verification by any third party entity. We also present the proof generation and verification performance for each circuit that is used to implement the MLP and CNN circuits. The individual circuits achieve fast and communication-light verification. We use the individual circuits to implement end-to-end watermark extraction and verification in \sys, however, these circuits can be combined to perform a myriad of tasks, including verifiable machine learning inference. 
\section{Conclusion}

This paper presented \sys, the first end-to-end watermark extraction and verification framework for DNNs based on zero-knowledge proofs. \sys utilizes zkSNARKs to enable a model owner to prove their right of ownership of a watermarked model while preserving privacy of watermark-sensitive data. We show \sys's end-to-end efficiency over multiple popular DNN benchmarks, and highlight the fact that our scheme is \textit{publically-verifiable}. Therefore, any third party can check the validity of the generated proofs in \sys. This work presents a paradigm shift from previous watermarking works by providing an end-to-end zero-knowledge approach to extracting watermarks, therefore allowing model owners to prove ownership of another model, without putting their original embedded watermarks at risk. 

\bibliographystyle{unsrt}
\bibliography{mybib}

\begin{thebibliography}{10}

\bibitem{chen2019looks}
Chaofan Chen, Oscar Li, Daniel Tao, Alina Barnett, Cynthia Rudin, and
  Jonathan~K Su.
\newblock This looks like that: deep learning for interpretable image
  recognition.
\newblock {\em Advances in neural information processing systems}, 32, 2019.

\bibitem{krizhevsky2017imagenet}
Alex Krizhevsky, Ilya Sutskever, and Geoffrey~E Hinton.
\newblock Imagenet classification with deep convolutional neural networks.
\newblock {\em Communications of the ACM}, 60(6):84--90, 2017.

\bibitem{hu2015face}
Guosheng Hu, Yongxin Yang, Dong Yi, Josef Kittler, William Christmas, Stan~Z
  Li, and Timothy Hospedales.
\newblock When face recognition meets with deep learning: an evaluation of
  convolutional neural networks for face recognition.
\newblock In {\em Proceedings of the IEEE international conference on computer
  vision workshops}, pages 142--150, 2015.

\bibitem{young2018recent}
Tom Young, Devamanyu Hazarika, Soujanya Poria, and Erik Cambria.
\newblock Recent trends in deep learning based natural language processing.
\newblock {\em ieee Computational intelligenCe magazine}, 13(3):55--75, 2018.

\bibitem{lin2021m6}
Junyang Lin, Rui Men, An~Yang, Chang Zhou, Yichang Zhang, Peng Wang, Jingren
  Zhou, Jie Tang, and Hongxia Yang.
\newblock M6: Multi-modality-to-multi-modality multitask mega-transformer for
  unified pretraining.
\newblock In {\em Proceedings of the 27th ACM SIGKDD Conference on Knowledge
  Discovery \& Data Mining}, pages 3251--3261, 2021.

\bibitem{shoeybi2019megatron}
Mohammad Shoeybi, Mostofa Patwary, Raul Puri, Patrick LeGresley, Jared Casper,
  and Bryan Catanzaro.
\newblock Megatron-lm: Training multi-billion parameter language models using
  model parallelism.
\newblock {\em arXiv preprint arXiv:1909.08053}, 2019.

\bibitem{darvish2019deepsigns}
Bita Darvish~Rouhani, Huili Chen, and Farinaz Koushanfar.
\newblock Deepsigns: An end-to-end watermarking framework for ownership
  protection of deep neural networks.
\newblock In {\em Proceedings of the Twenty-Fourth International Conference on
  Architectural Support for Programming Languages and Operating Systems}, pages
  485--497, 2019.

\bibitem{uchida2017embedding}
Yusuke Uchida, Yuki Nagai, Shigeyuki Sakazawa, and Shin'ichi Satoh.
\newblock Embedding watermarks into deep neural networks.
\newblock In {\em Proceedings of the 2017 ACM on international conference on
  multimedia retrieval}, pages 269--277, 2017.

\bibitem{chen2018deepmarks}
Huili Chen, Bita~Darvish Rohani, and Farinaz Koushanfar.
\newblock Deepmarks: A digital fingerprinting framework for deep neural
  networks.
\newblock {\em arXiv preprint arXiv:1804.03648}, 2018.

\bibitem{bitan2022using}
Dor Bitan, Ran Canetti, Shafi Goldwasser, and Rebecca Wexler.
\newblock Using zero-knowledge to reconcile law enforcement secrecy and fair
  trial rights in criminal cases.
\newblock {\em Available at SSRN}, 2022.

\bibitem{bamberger2022verification}
Kenneth~A Bamberger, Ran Canetti, Shafi Goldwasser, Rebecca Wexler, and Evan~J
  Zimmerman.
\newblock Verification dilemmas in law and the promise of zero-knowledge
  proofs.
\newblock {\em Berkeley Technology Law Journal}, 37(1), 2022.

\bibitem{blum1991noninteractive}
Manuel Blum, Alfredo De~Santis, Silvio Micali, and Giuseppe Persiano.
\newblock Noninteractive zero-knowledge.
\newblock {\em SIAM Journal on Computing}, 20(6):1084--1118, 1991.

\bibitem{Li2021ASO}
Yue Li, Hongxia Wang, and Mauro Barni.
\newblock A survey of deep neural network watermarking techniques.
\newblock {\em Neurocomputing}, 461:171--193, 2021.

\bibitem{nagai2018digital}
Yuki Nagai, Yusuke Uchida, Shigeyuki Sakazawa, and Shin’ichi Satoh.
\newblock Digital watermarking for deep neural networks.
\newblock {\em International Journal of Multimedia Information Retrieval},
  7(1):3--16, 2018.

\bibitem{adi2018turning}
Yossi Adi, Carsten Baum, Moustapha Cisse, Benny Pinkas, and Joseph Keshet.
\newblock Turning your weakness into a strength: Watermarking deep neural
  networks by backdooring.
\newblock In {\em 27th USENIX Security Symposium (USENIX Security 18)}, pages
  1615--1631, 2018.

\bibitem{Zhang2020ModelWF}
Jie Zhang, Dongdong Chen, Jing Liao, Han Fang, Weiming Zhang, Wenbo Zhou, Hao
  Cui, and Nenghai Yu.
\newblock Model watermarking for image processing networks.
\newblock In {\em Proceedings of the AAAI Conference on Artificial
  Intelligence}, volume~34, pages 12805--12812, 2020.

\bibitem{ben2019aurora}
Eli Ben-Sasson, Alessandro Chiesa, Michael Riabzev, Nicholas Spooner, Madars
  Virza, and Nicholas~P Ward.
\newblock Aurora: Transparent succinct arguments for r1cs.
\newblock In {\em Annual international conference on the theory and
  applications of cryptographic techniques}, pages 103--128. Springer, 2019.

\bibitem{danezis2014square}
George Danezis, C{\'e}dric Fournet, Jens Groth, and Markulf Kohlweiss.
\newblock Square span programs with applications to succinct nizk arguments.
\newblock In {\em International Conference on the Theory and Application of
  Cryptology and Information Security}, pages 532--550. Springer, 2014.

\bibitem{sasson2014zerocash}
Eli~Ben Sasson, Alessandro Chiesa, Christina Garman, Matthew Green, Ian Miers,
  Eran Tromer, and Madars Virza.
\newblock Zerocash: Decentralized anonymous payments from bitcoin.
\newblock In {\em 2014 IEEE symposium on security and privacy}, pages 459--474.
  IEEE, 2014.

\bibitem{maller2019sonic}
Mary Maller, Sean Bowe, Markulf Kohlweiss, and Sarah Meiklejohn.
\newblock Sonic: Zero-knowledge snarks from linear-size universal and updatable
  structured reference strings.
\newblock In {\em Proceedings of the 2019 ACM SIGSAC Conference on Computer and
  Communications Security}, pages 2111--2128, 2019.

\bibitem{groth2016size}
Jens Groth.
\newblock On the size of pairing-based non-interactive arguments.
\newblock In {\em Annual international conference on the theory and
  applications of cryptographic techniques}, pages 305--326. Springer, 2016.

\bibitem{xjsnark}
{xJsnark}.
\newblock \url{https://github.com/akosba/xjsnark}.

\bibitem{libsnark}
{libsnark}.
\newblock \url{https://github.com/scipr-lab/libsnark}.

\bibitem{garg2022succinct}
Sanjam Garg, Abhishek Jain, Zhengzhong Jin, and Yinuo Zhang.
\newblock Succinct zero knowledge for floating point computations.
\newblock In {\em Proceedings of the 2022 ACM SIGSAC Conference on Computer and
  Communications Security}, pages 1203--1216, 2022.

\bibitem{weng2021mystique}
Chenkai Weng, Kang Yang, Xiang Xie, Jonathan Katz, and Xiao Wang.
\newblock Mystique: Efficient conversions for $\{$Zero-Knowledge$\}$ proofs
  with applications to machine learning.
\newblock In {\em 30th USENIX Security Symposium (USENIX Security 21)}, pages
  501--518, 2021.

\bibitem{wan2022zk}
Zhiguo Wan, Yan Zhou, and Kui Ren.
\newblock zk-authfeed: Protecting data feed to smart contracts with
  authenticated zero knowledge proof.
\newblock {\em IEEE Transactions on Dependable and Secure Computing}, 2022.

\end{thebibliography}

\end{document}